\documentclass[aps,pra,twocolumn,showpacs,10pt,floatfix]{revtex4}
\usepackage{graphicx}
\usepackage{amssymb}
\usepackage{amsmath}

\begin{document}
\title{Electromagnetic induced transparency and slow light in interacting quantum degenerate atomic gases}
\author{H. H. Jen$^{1,*}$, Bo Xiong$^{1}$, Ite A. Yu$^{1,2}$, and Daw-Wei Wang$^{1,2,3}$}
\affiliation{$^{1}$Physics Department, National Tsing Hua University, Hsinchu, Taiwan, R. O. C.\\
$^{2}$Frontier Research Center on Fundamental and Applied Sciences of Matter, National Tsing Hua University, Hsinchu, Taiwan, R. O. C.\\
$^{3}$Physics Division, National Center for Theoretical Sciences, Hsinchu, Taiwan, R. O. C.}

\newcommand{\p}{\mathbf{p}}
\newcommand{\q}{\mathbf{q}}
\date{\today}
\email{sappyjen@gmail.com}
\begin{abstract}
We systematically develop the full quantum theory for the electromagnetic induced transparency (EIT) and slow light properties in ultracold Bose and Fermi gases. It shows a very different property from the classical theory which assumes frozen atomic motion. For example, the speed of light inside the atomic gases can be changed dramatically near the Bose-Einstein condensation temperature, while the presence of the Fermi sea can destroy the EIT effect even at zero temperature. From experimental point of view, such quantum EIT property is mostly manifested in the counter-propagating excitation schemes in either the low-lying Rydberg transition with a narrow line width or in the D2 transitions with a very weak coupling field. We further investigate the interaction effects on the EIT for a weakly interacting Bose-Einstein condensate, showing an inhomogeneous broadening of the EIT profile and nontrivial change of the light speed due to the quantum many-body effects beyond mean field energy shifts.
\end{abstract}
\maketitle

\section{Introduction}
Electromagnetic induced transparency (EIT) has been extensively studied in quantum optics \cite{Lukin_rmp,Fleischhauer_rmp} and applied to serve as quantum memory elements through light-atom interactions \cite{QInterface}.  It is also realized in ultracold Bose gas \cite{BEC_Hau} to slow down the light speed in an unprecedented way.  The transparency along with significant group velocity delay opens up great opportunities of light storage and retrieval in neutral atomic systems, as one of the candidates in quantum information processing.

In the classical theory of EIT, it usually assumes frozen atomic motion which is valid only in limited conditions when the recoil momenta from laser interactions are negligible.\ When the thermal cloud at several hundreds $\mu$K or higher is considered, the frequency shift resulted from the atomic motion is comparable to the EIT transparency window.\ The Doppler effect can cause decoherence to the probe field due to the large spread of atomic velocity distribution \cite{doppler}.\ In the other regime of ultracold atoms considered in this article, a Gross-Pitaevskii (GP) equation has been used in Bose-Einstein condensates (BEC) in the zero temperature limit to investigate light propagation dynamics \cite{gp}.\ The dispersion relations of light are unaffected due to the extremely low energy atomic dynamics ($\lesssim$ kHz) even in the counter-propagating scheme, providing the great opportunity and experimental flexibility in
the quantum information application based on the EIT effect.\ However,this is not the case for example when Rydberg states are considered in the application \cite{Rydberg_rev}, in which the recoil energy of light imparting on atoms is comparable to the inverse of lifetime for Rydberg transitions.\ Also at finite temperature across the quantum degeneracy, atomic Bose or Fermi statistics starts to play a crucial role in determining EIT property where GP equation is not applicable.\ We note that here the classical theory refers to the conventional quantum optical treatment in which atomic motions are assumed frozen, and the quantum theory in comparison refers to the generalization of the quantum statistical nature of atoms. For both theories, the optical field is treated classically.

To develop the general quantum theory of EIT, in this paper we include the atomic kinetic energy explicitly to get the dark state dynamics. Compared to the energy scale of the coupling field and the decay rate of the excited state, the recoil kinetic energy is found to be significant mostly in the counter-propagating scheme either in the low-lying Rydberg transition or the D2 transition of a very weak field. For bosonic gases, the group velocity of the transmitted light is strongly modified near the BEC transition temperature. For fermionic gases, on the other hand, there is no EIT allowed due to the presence of a Fermi sea. We also calculate the EIT profile for co-propagating scheme, and find almost no effects of such quantum statistics, and EIT is robust in a wide range of temperature (even across the quantum degenerate temperature) for both bosonic and fermionic atoms. Finally, we also include interaction effects in bosonic gases. We find that, besides of the simple mean-field energy shift, the EIT profile can be broadened by the many-body effects, significantly increase the group velocity when the interaction is stronger. Our results show that the EIT experiment can be a good candidate to explore the quantum many-body properties of the underlying quantum gases.

In this article, we review the classical theory on EIT in Sec. II and derive the general quantum theory that includes atomic dynamics and its quantum statistics in alkali atomic gases in Sec. III.\ We demonstrate the EIT property in noninteracting Bose and Fermi gases using ultraviolet transitions of low-lying Rydberg states and D2 transitions for both co- and counter-propagating excitation schemes.\ In Sec. IV, we apply the mean field approach to the EIT in a weakly interacting BEC and conclude in Sec. V.

\section{Classical theory on EIT}

\begin{figure}[t]
\centering\includegraphics[height=6.5cm, width=8.5cm]{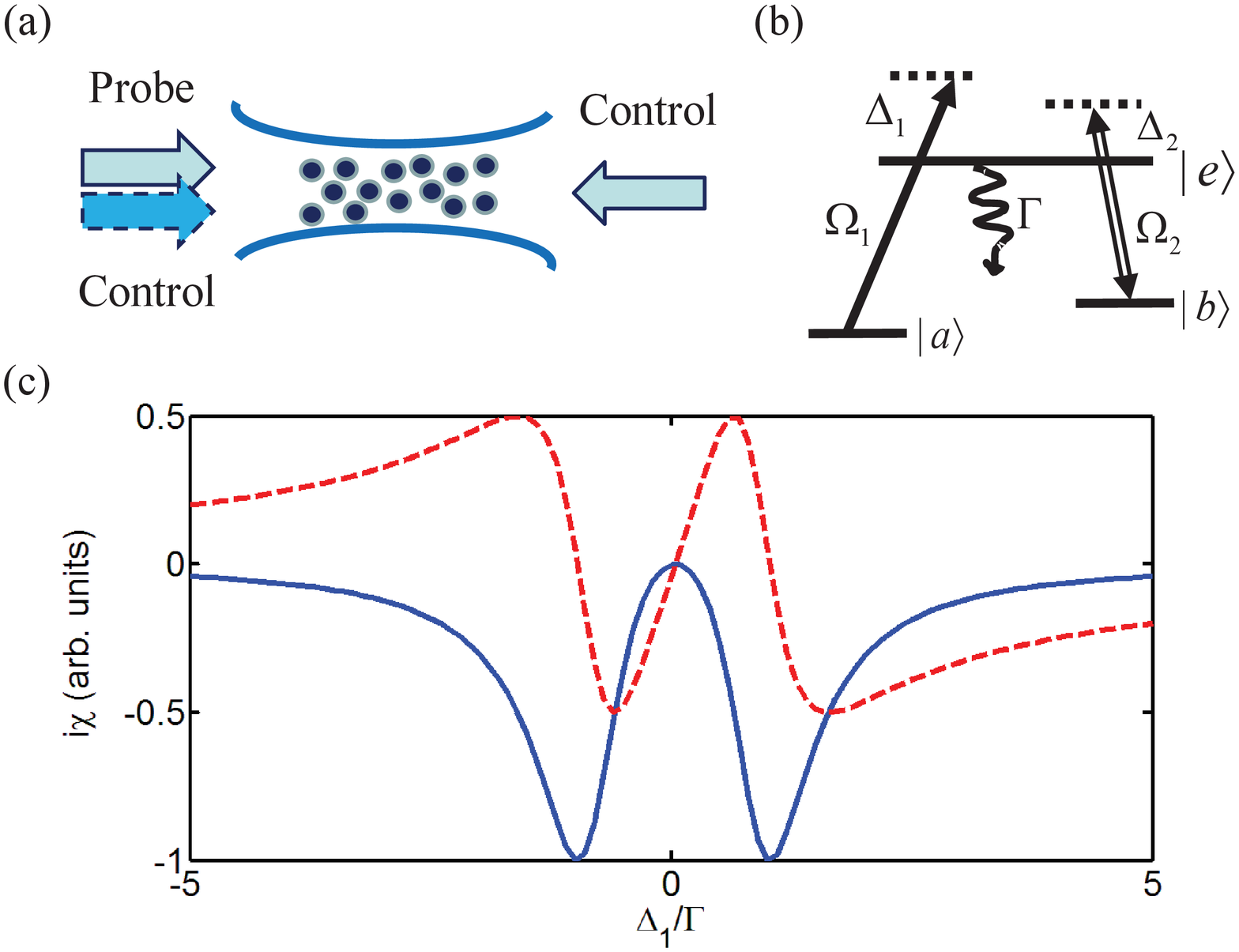}
\caption{(Color online) Electromagnetic induced transparency (EIT) in $\Lambda$-type atoms.\ (a) The atoms are interacting with two laser fields either through co (dashed)- or counter (solid)-propagating scheme.\ (b) Control $\Omega_{2}$ and probe fields $\Omega_{1}$ couple two hyperfine ground states $|a\rangle$ and $|b\rangle$ with the excited state $|e\rangle$ with detunings $\Delta_2$ and $\Delta_1$ respectively.\ $\Gamma$ is the spontaneous emission rate of $|e\rangle$.\ (c) Linear susceptibility $\chi$ times $i$ versus probe light detuning $\Delta_1$ in unit of $\Gamma$.\ Absorption (solid-blue) and dispersion (dashed-red) profiles are plotted from the classical theory of EIT in resonance of the control field $\Delta_2=0$.}%
\label{eit}
\end{figure}

The classical theory assumes frozen atomic motion so that the physics of EIT mainly relies on the quantum interference of two laser fields coupled with $\Lambda$ type atomic structure as shown in Fig. $\ref{eit}$.\ The atoms interact with control and probe fields for transitions $|b\rangle\rightarrow|e\rangle$ and $|a\rangle\rightarrow|e\rangle$ of Rabi frequencies $\Omega_{2}$ and $\Omega_{1}$ respectively, either in co- or counter-propagating excitation scheme.\ Single photon detunings are defined as $\Delta_{1}=\omega_{1}-(\omega_{e}-\omega_{a}),$ $\Delta_{2}=\omega_{2}-(\omega_{e}-\omega_{b})$.\ The relatively weak probe field $\Omega_1$ propagates into a medium interacting with a strong control field $\Omega_2$, which modifies the energy of the excited state via AC Stark shift.\ The one dimensional wave equation for slow-varying probe electric field $E(z,t)$ can be described by \cite{QO}% 

\begin{equation}
\frac{\partial E(z,t)}{\partial z}+\frac{1}{c}\frac{\partial E(z,t)}{\partial t}=\frac{ik_{1}}{2\epsilon_{0}}P(z,t),
\end{equation}
where $\epsilon_0$ is the vacuum permittivity and the atomic polarization in Fourier space is $P(z,\omega)=\chi(z,\omega)E(z,\omega)$ with the electric susceptibility $\chi(z,\omega)$.\ Along with the optical Bloch equations of atoms \cite{QO}, the above makes up a self-consistent Maxwell-Bloch equation which fully accounts for atomic coherence and light propagation dynamics.\ Use $P(z,t)=\left\langle \rho d_{ae}\tilde{\sigma}_{ae}(z,t)\right\rangle$ with the atomic density $\rho$, dipole matrix element $d_{ae}$, and slow-varying atomic coherence operator $\tilde{\sigma}_{ae}(z,t)$ in a rotating frame of probe light central frequency $\omega_1$ and wavevector $k_1$, we may express susceptibility and group velocity in resonance of the constant control field ($\Delta_2=0$) as \cite{Lukin_rmp}

\begin{eqnarray}
\chi(\omega_1)&=&\frac{\rho d^2_{ae}}{\hbar}\frac{\Delta_1}{\Omega_2^2-(\Delta_1+i\Gamma)\Delta_1},\nonumber\\
v_{g}&=&\frac{c}{1+N|g|^{2}\left\{\frac{\left(\Omega_2^2+\Delta_1^2\right)\left[\left(\Omega_2^2-\Delta_1^2\right)^2-\Delta_1^2\Gamma^2\right]}{\left[\left(\Omega_2^2-\Delta_1^2\right)^2+\Delta_1^2\Gamma^2\right]^2}\right\}},\nonumber\\&=&\frac{c}{1+N|g|^2/\Omega_2^2}\Big |_{\Delta_1=0},\label{classical}
\end{eqnarray}
where $g=d_{ae}\sqrt{\omega_1/(2\hbar\epsilon_0V)}$ is field coupling constant with the quantization volume $V$.\ $N|g|^{2}$ associates with optical depth (OD) that determines the reduction of group velocity and $N$ is the total number of atoms.\ The group velocity is derived from $v_{g}=d\omega_{1}/dk(\omega_{1})$ where $k(\omega_{1})=\omega_{1}/c+\operatorname{Im}\left[ik_{1}\chi(\omega_1)/(2\epsilon_{0})\right]$.\ In the above, we specifically show the results at the zero absorption pint where $\Delta_1=0$.\ The absorption and dispersion profiles are shown in Fig. 1(c) for the Rabi frequency $\Omega_2=\Gamma$.\ The typical results from classical theory show that the width of transparency window is determined by $2\Omega_2$ with a normal dispersion in it.\ The group velocity at zero absorption point depends on the control field strength and optical depth, which decreases as $\Omega_2$ decreases or with a larger OD.\ Another useful property of oscillator strength, $f=2m_e\omega d^2/(3\hbar e^2)$ \cite{Rydberg}, will be used later to characterize the coupling strength between S and P level transitions of frequency $\omega$ and dipole matrix element $d$, and $m_e$ is the mass of electron with charge $e$.

The above results delineate the essence of induced transparency through electromagnetic couplings from classical theory.\ In the next section, we generalize the treatment of EIT to include the atomic dynamics and quantum statistics especially in noninteracting quantum degenerate gases at finite temperature. 

\section{Noninteracting quantum degenerate gases at finite temperature}

In general, the Hamiltonian ($H=H_{A}+H_{AL}$) of noninteracting quantum degenerate gases in momentum space is 
\begin{eqnarray}
H_{A}  & =&\sum_{\p}\frac{\p^{2}}{2m}\left(  \hat{a}_{\p}^{\dagger}\hat{a}_{\p}+\hat{b}_{\p}^{\dagger}\hat{b}_{\p}+\hat{e}_{\p}^{\dagger}\hat{e}_{\p}\right)
+\hbar\Delta_{1}\sum_{\p}\hat{a}_{\p}^{\dagger}\hat{a}_{\p} \nonumber\\
&+&\hbar\Delta_{2}\sum_{\p}\hat{b}_{\p}^{\dagger}\hat{b}_{\p}\text{,}\nonumber\\
H_{AL}  & =&-\hbar\Omega_{1}\sum_{\p}\hat{e}_{\p+\p_{1}}^{\dagger}\hat{a}_{\p}-\hbar\Omega_{2}\sum_{\p}\hat{e}_{\p+\p_{2}}^{\dagger}\hat{b}_{\p}+h.c.,\label{nonH}
\end{eqnarray}
where $H_{A}$ includes the kinetic and internal state energy for atoms, and $H_{AL}$ represents the atom-light interactions.\ Dipole approximation, $-\mathbf{d}\cdot\mathbf{E},$ and rotating wave approximation (RWA) have been made to the atom-light interactions.\ $\p_{1,2}$ are recoil momenta of the probe and control fields respectively.\ Note that we have used interaction picture for atom and light fields to remove the free evolution terms (we keep the atomic kinetic energy), and atom fields satisfy the bosonic or fermionic commutation relations, e.g. $[\hat{a}_{\p},\hat{a}_{\p'}^{\dagger}]_{\pm}=\delta_{\p\p'}$.

We may express the Hamiltonian in terms of the basis, $\mathbf{A_\p}=%
\begin{bmatrix}
\hat{a}_{\p},&\hat{b}_{\p+\p_r},&\hat{e}_{\p+\p_{1}}%
\end{bmatrix}^{T}$, with $\p_r\equiv \p_{1}-\p_{2}$, and let $H=\sum_{\p}\mathbf{A}^{\dag}_\p\cdot\hat{M}_\p\cdot\mathbf{A_\p}$, where the matrix $\hat{M}_\p$ is ($\hbar=1$)
\begin{equation}
\hat{M_\p}=%
\begin{bmatrix}
\Delta_{1}+\frac{\p^{2}}{2m} & 0 & -\Omega_{1}\\
0 & \Delta_{2}+\frac{\left(  \p+\p_r\right)  ^{2}}{2m} & -\Omega_{2}\\
-\Omega_{1} & -\Omega_{2} & \frac{\left(  \p+\p_{1}\right)  ^{2}}{2m}%
\end{bmatrix}.\label{M}
\end{equation}
In general, we can diagonalize $H$ by a similarity transformation $\hat{S}_\p$ such that in the new basis $\mathbf{B}_\p=\hat{S}_\p\mathbf{A}_\p$, it becomes a diagonal matrix $\hat{\Lambda}_\p=\hat{S}_\p\hat{M}_\p\hat{S}^{-1}_\p$.\ We are interested in the regime when the probe field is weak so we can solve the above Hamiltonian in the leading order of small $\Omega_{1}$.\ Under the adiabatic approximation, the atoms initially prepared on the ground state $|a\rangle$ follow these light fields and form a dark state, $\hat{\beta}_\p^\dag$ which is immune to spontaneous emission of the excited state (up to a normalization constant),
\begin{widetext}
\begin{eqnarray}
&&\hat{\beta}_{\p}^{\dag}|0\rangle=\Bigg[ \hat{a}_{\p}^{\dagger}+ \Omega_{1}\cos\phi_\p\sin\phi_\p\Bigg(\frac{1}{\epsilon_{D}(\p)-\epsilon_{B_+}(\p)} -\frac{1}{\epsilon_{D}(\p)-\epsilon_{B_-}(\p)}\Bigg)\hat{b}_{\p+\p_r}^{\dagger} \nonumber\\
&&-\Omega_{1}\Bigg(\frac{\sin^2\phi_\p}{\epsilon_{D}(\p)-\epsilon_{B_+}(\p)}+\frac{\cos^2\phi_\p}{\epsilon_{D}(\p)-\epsilon_{B_-}(\p)}\Bigg)\hat{e}_{\p+\p_{1}}^{\dagger}\Bigg] |0\rangle,\label{dark}
\end{eqnarray}
\end{widetext}
where $\epsilon_D(\p)$ and $\epsilon_{B_\pm}(\p)$ are the eigenvalues of the unperturbed Hamiltonian for the dark and two bright states respectively,
\begin{eqnarray}
\epsilon_{D}(\p)&=&\Delta_{1}+\frac{\p^{2}}{2m},\nonumber\\
\epsilon_{B_\pm}(\p)&=&\frac{\bar{\Delta}_{2}+\frac{\left(\p+\p_{1}\right)  ^{2}%
}{2m}\pm\sqrt{\left[\bar{\Delta}_{2}-\frac{\left(\p+\p_{1}\right)  ^{2}}{2m}\right]^{2}+4\Omega_{2}^{2}}}{2},\nonumber\\
\end{eqnarray}
where $\bar{\Delta}_{2}=\Delta_{2}+\left(\p+\p_r\right)^{2}/(2m)$ and 
\begin{equation}
\cos\phi_\p=\sqrt{\frac{\epsilon_{B_+}(\p)-\frac{\left(\p+\p_{1}\right)^{2}}{2m}}{\epsilon_{B_+}(\p)-\epsilon_{B_-}(\p)}}.
\end{equation}
The angle $\phi_\p$ represents the mixing of the hyperfine ground and excited states coupled by the control field.\  

This dark state provides the ingredient to the dispersion and absorption of light propagation in the atoms.\ For a noninteracting quantum degenerate gas, we derive the linear susceptibility $\chi(\omega_1)$ under the dark state basis assuming a constant control field,
\begin{eqnarray}
\chi(\omega_1)&\equiv&\frac{d^2_{ae}}{\hbar\Omega_1A}\int d^2\mathbf{r}_\perp \left\langle\hat{\Psi}_a^\dagger(\mathbf{r})\hat{\Psi}_e(\mathbf{r})\right\rangle e^{-i\p_{1}\cdot\mathbf{r}/\hbar}\nonumber\\
&=&\frac{\rho d^2_{ae}}{N\hbar\Omega_1}\sum_{\p,\p'}\left\langle\hat{a}_{\p}^{\dag}\hat{e}_{\p^{\prime}}\right\rangle \frac{1}{A}\int d^2\mathbf{r}_\perp e^{i(\p^{\prime}-\p-\p_{1})\cdot\mathbf{r}/\hbar},\nonumber\\
&=&\frac{\rho d^2_{ae}}{\hbar N}\sum_\p F_\p n_{\beta_\p},\label{quantum}
\end{eqnarray} 
where
\begin{widetext}
\begin{eqnarray}
F_\p &=& -\left[\frac{\sin^{2}\phi_\p}{\epsilon_{D}(\p)-\epsilon_{B_+}(\p)}+\frac{\cos^{2}\phi_\p}{\epsilon_{D}(\p)-\epsilon_{B_-}(\p)}\right],\nonumber\\
&=&\frac{\Delta_1-\Delta_2+\frac{\p^{2}}{2m}-\frac{\left(\p+\p_{r}\right)^{2}}{2m}%
}{\Omega_{2}^{2}-\left[\Delta_{1}+\frac{\p^{2}}{2m}-\frac{\left(\p+\p_{1}\right)^{2}}{2m}+i\Gamma\right]\left[\Delta_1-\Delta_2+\frac{\p^{2}}%
{2m}-\frac{\left(\p+\p_{r}\right)^{2}}{2m}\right]}.\label{kernel}
\label{nonint}
\end{eqnarray}
\end{widetext}

$\Psi(\mathbf{r})$ is the atomic field operator in real space and $\mathbf{r}_\perp$ is the direction perpendicular to $\p_1$ with the cross section area $A$.\ The expectation value of Eq. (\ref{quantum}) is calculated from the dark state in Eq. (\ref{dark}).\ Note that $\chi(\omega_1)$ has spatial dependence of $z$ which is implicit because we assume a uniform atomic density $\rho$.\ The boson/fermion number distribution is $n_{\beta_{\p}}=1/[\exp\{[\epsilon_{D}(\p)-\mu]/(k_BT)\}\mp 1]$ where $k_B$ is the Boltzmann constant, and the chemical potential $\mu$ is determined by the number conservation $\sum_{\p}n_{\beta_{\p}}=N$.\ For convenience we define a kernel function, $F_\p$ which will be encountered quite often throughout this paper.\ The second line of $F_\p$ is derived by adding a phenomenonological spontaneous decay rate ($\Gamma$) of the excited state into the noninteracting Hamiltonian.\ The dispersion relation from the above quantum theoretical results indicates a frequency shift for single and two-photon detunings from atomic central ($\p$) and recoil ($\p_1$ and $\p_r$) momentum distributions.\  The absorption width of probe field is mainly determined by the energy scale of natural decay rate $\Gamma$.\ Note that Eq. (\ref{kernel}) reduces to the classical results when the atomic motion is assumed frozen.\ In subsequent subsections, we study noninteracting bosonic/fermionic gases and investigate the EIT property when the excited states are low-lying Rydberg state and D2 transitions for both co- and counter-propagating excitation schemes.

\subsection{Results of counterpropagating excitation scheme}

From Eq. (\ref{kernel}), there are three energy scales, $\Omega_2$, $\Gamma$, and $E_r=\p_r^2/(2m)$, that play the major roles in the dispersion and absorption profiles of EIT.\ When D1 or D2 transition of the alkali atoms is considered, the spontaneous decay rate ($\Gamma$) is in the range of MHz which exceeds low energy atomic dynamics ($\lesssim$ kHz).\ The quantum theory of EIT in the previous section reduces to the classical results as if the atomic motion is frozen when the control field strength $\Omega_2$ is also in the order of $\Gamma$.\ However, the spontaneous emission rate of the Rydberg state can be in the range of kHz \cite{Rydberg} and even below for larger principal quantum number \cite{Rydberg_rev}.\ This motivates us to use Rydberg transitions to investigate the quantum effects of EIT where $\Omega_2\approx E_r\approx\Gamma$.\ For comparison, we also calculate D2 transitions when using a weak control field, $\Omega_2\approx E_r\ll\Gamma$, which can show quite different properties compared to classical results.

\subsubsection{Low-lying Rydberg bosons}

Here we first consider the counterpropagating excitation scheme as shown in Fig. \ref{eit}(a) and choose the excited levels of rubidium (Rb) and potassium (K) as low-lying Rydberg states.\ The radiative lifetimes for low-lying Rydberg atoms ($n$P levels) \cite{Rydberg} can reach to the comparable recoil energy scale when the principal quantum number is around $n=24$ \cite{Gounand}.\ Note that the dipole-dipole interaction of Rydberg atoms can be significant and relevant here.\ To simplify the discussion in this work, we only consider the situation that the dipole-dipole interaction or dipole blockade is weak, e.g. low-lying Rydberg excitations ($n<30$) \cite{30P}.\ The long-range van der Waals interaction coefficient $|\text{C}_6|$ for $30$P-$30$P rubidium asymptotes is in the range of $(0.26-4.7)\times 10^{16}$ a.u. \cite{C6} that gives us the blockade radius $\text{R}_\text{c}=0.88-1.29$ $\mu$m \cite{superatom,superatom2} for a density $\rho=10^{14}$ $\text{cm}^{-3}$ with the effective excitation Rabi frequency $\Omega=2\pi\times100$ kHz.\ This radius provides an upper estimation of the strength of dipole blockade.\ For potassium atoms ($n=30$), we have blockade radius $\text{R}_\text{c}=1.3-1.67$ $\mu$m if the same atomic density and excitations are applied.\ The dipole-dipole interaction energy can be further reduced by decreasing $\Omega$ so that it is not relevant when low-lying Rydberg states are concerned.\ In the future, we can also include the dipole-dipole interaction of Rydberg atoms. 
 
For a noninteracting Bose gas, the transition temperature can be found analytically for a three dimensional uniform gas $T_c=2\pi\hbar^2(\rho/\zeta(3/2))^{(2/3)}/(mk_B)$ \cite{BEC} where $\zeta$ is Riemann zeta function, and $\rho$ is atomic density of mass $m$.\ The chemical potential approaches the minimum of eigenenergy $\epsilon_D(\p=0)=\Delta_{1}$ when temperature approaches $T_{c}$.\ We find $T_c\approx 0.4~\mu$K for $^{87}$Rb of a density $\rho=10^{14}~\text{cm}^{-3}$.\ Throughout the article we use this atomic density for demonstration of the results.

At $T<T_{c},$ the condensate particle density ($\rho_{c}$) is derived from subtraction of the excited particle density $\rho_{c}(T)=\rho-\rho_{ex}$.\ The electric susceptibility of finite temperature are resulted from the condensate and non condensate parts,
\begin{equation}
\chi(\omega_1,T)=\frac{\rho d^2_{ae}}{\hbar}\left[\frac{\rho_{c}(T)}{\rho}F_{\p=0} +\frac{1}{N}\sum_{\p\neq0}F_\p n_{\beta_{\p}}\right],\label{bec_coh}
\end{equation}
where the first term is exactly the classical result in Eq. (\ref{classical}) if $\rho_c=\rho$, and the second term is thermal depletion.\
To calculate the group velocity for the probe light, we need the oscillator strengths ($f$) between the ground and low-lying Rydberg states, $|21\text{P}\rangle$ of K and $|24\text{P}\rangle$ of Rb, which are $3.1\times 10^{-7}$ and $2.8\times 10^{-6}$ respectively \cite{oscillator}.  The branch ratio of transitions to $P_{1/2}$ are large enough that $P_{3/2}/P_{1/2}\approx 4$ and $8$ for K and Rb respectively \cite{branch1,branch2} so we may take the values of oscillator strengths as a correct order of magnitude.\ For D$_2$ transitions of potassium and rubidium atoms, $f=0.658$ and $0.682$, much larger than the low-lying Rydberg states.\

\begin{figure}[t]
\centering\includegraphics[height=4.6cm, width=8.5cm]{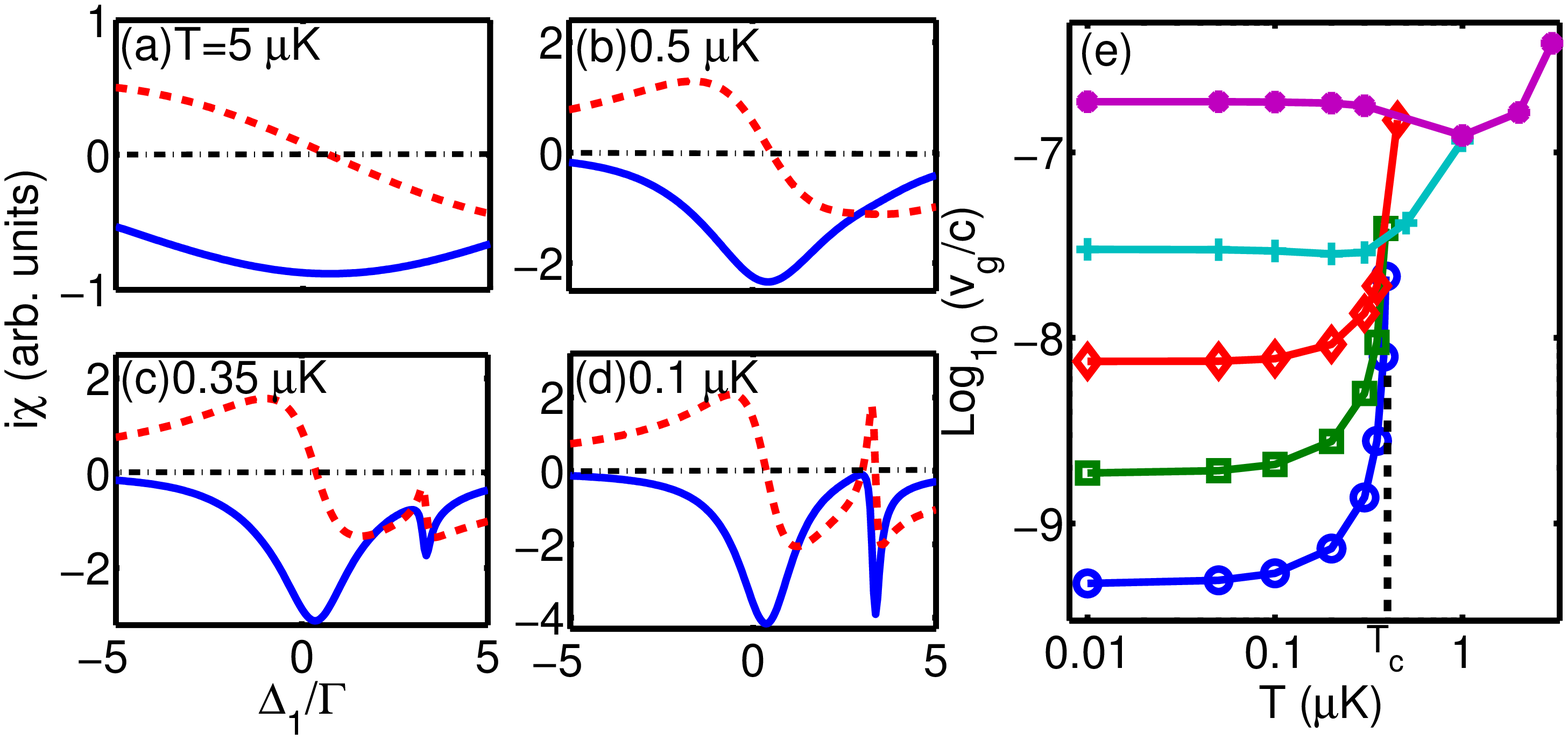}%[height=2.1in, width=3.60in]
\caption{(Color online) Counterpropagating excitation scheme for EIT in bosons using the Rydberg transition ($\Gamma^{-1}=28.3\mu\text{s}$) of $^{87}$Rb.\ The control field is $\Omega_2=1\Gamma$ with the detuning $\Delta_2=0$ which is used throughout the article if not specified.\ Absorption (Re[$i\chi$])(solid-blue) and dispersion (Im[$i\chi$])(dash-red) profiles are plotted at different temperatures (a) T=$5$, (b) $0.5$, (c) $0.35$, and (d) $0.1\mu$K where transition temperature $T_c\approx 0.4\mu$K (straight dotted line in (e)).\ Dash-dotted black line guides the eye to the zero of the plots.\ In (e), the group velocity $v_g$ is plotted for various control fields, $\Omega_2/\Gamma=0.5(\bigcirc)$, $1(\Box)$, $2(\Diamond)$, $4(+)$, $10(\bullet)$ over temperatures in log scale.\ The detuning is set as $\Delta_2=-\p_r^2/(2m)$ so $v_g$ is calculated at the transparency condition of $\Delta_1=0$.\ The solid lines guide the eyes for connecting data points.}%
\label{counter_B}
\end{figure}

In Fig. \ref{counter_B}, we show the electric susceptibility ($\chi$) and group velocity ($v_g$) of the probe field in $^{87}$Rb.\ The excited state is chosen as low-lying Rydberg states $|e\rangle=|24\text{P}_{3/2}\rangle$ with spontaneous decay time $\Gamma^{-1}=28.3$ $\mu$s.\ When $T>T_c$ in (a) and (b), there is no transparency due to the large bosonic momentum distribution that deviates from the resonance condition of free absorption.\ The relaxation due to the Doppler width at high temperature can be estimated as $k_BT\text{ln}2/\hbar$.\ Just below $T_c$ in (c), two absorption (Re[$i\chi$]) lines appear, and the transparency condition starts to emerge at $\Delta_1\approx 3\Gamma$ that is due to the recoil energy shift $(E_r)$ for the probe field wavelength $299$ nm.\ When $T$ is much smaller than $T_c$ in (d), the condensate component dominates over the thermal one, and we have a perfect EIT with zero absorption.\ The two absorption peaks can be identified as $\Delta_1=\{E_1+E_r\pm[(E_1-E_r)^2+4\hbar^2\Omega_2^2]^{1/2}\}/(2\hbar)$ where $E_1=\p_1^2/(2m)$.\ Note that the absorption peak in plot (a) when $T\gg T_c$ can be identified as $\Delta_1=E_1/\hbar$ due to the thermally-averaged transparency condition ($\Delta_1=E_r/\hbar$).

In Fig. \ref{counter_B}(e), the group velocity is plotted for various control fields from $0.5$ to $10\Gamma$.\
For small $\Omega_2$, we find a significant reduction of group velocity when $T<T_c$, and it saturates in the limit of $T\rightarrow 0$.\ There is no slow light at a higher temperature because no transparency is supported if $\Omega_2$ is too small.\ This contrasts to the conventional experiments using D1 or D2 transitions of Rb where $\Omega_2$ is large \cite{Lukin_rmp,Fleischhauer_rmp}.\ From Fig. \ref{counter_B}(e), we demonstrate the interplay between the control fields $\Omega_2$ and temperature $T$ for the effect of EIT on slow light propagation.\ We find that this may be an efficient and precise method to determine the $T_c$ in bosonic quantum gases. 

\subsubsection{D2 transition bosons}

\begin{figure}[t]
\centering\includegraphics[height=4.6cm, width=8.5cm]{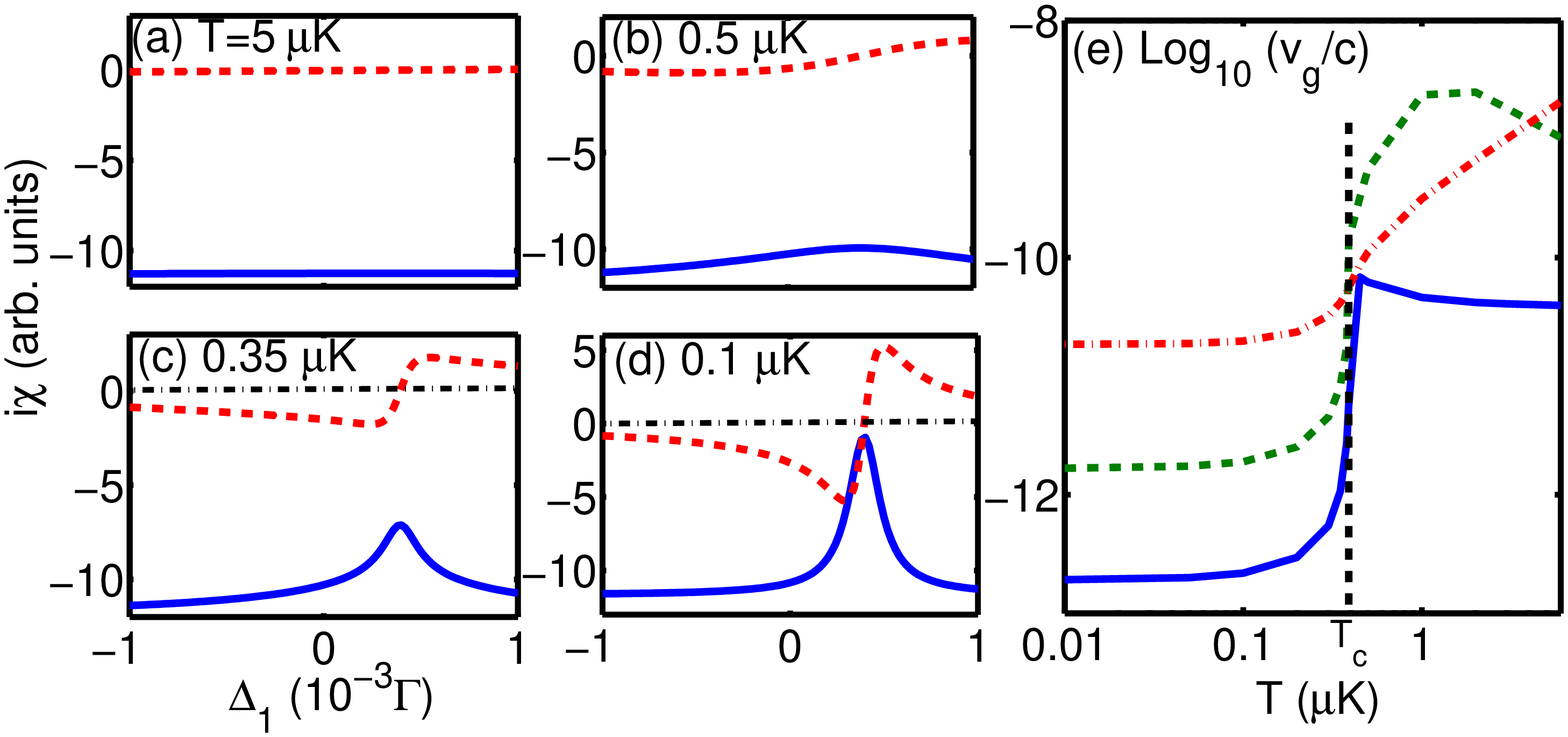}%[height=2.1in, width=3.60in]
\caption{(Color online) Counterpropagating excitation scheme for EIT in bosons using D2 transition ($\Gamma^{-1}=26$ns) of $^{87}$Rb.\ Absorption (Re[$i\chi$])(solid-blue) and dispersion (Im[$i\chi$])(dash-red) profiles are plotted as a comparison to Fig. \ref{counter_B} at the temperatures (a) T=$5$, (b) $0.5$, (c) $0.35$, and (d) $0.1\mu$K.\ Dash-dotted black line guides the eye to the zero of the plots, and $\Omega_2=0.01\Gamma$, $\Delta_2=0$.\ In (e), the group velocity $v_g$ is plotted for $\Omega_2/\Gamma=0.003(\text{solid-blue})$, $0.01(\text{dashed-green})$, $0.03(\text{dash-dotted red})$ over temperatures in log scale across $T_c$ (straight dotted line).}%
\label{D2}
\end{figure}

In Fig. \ref{D2}, we demonstrate the similar results compared to Fig. \ref{counter_B} when the D2 transition ($\Gamma^{-1}=26$ ns) is chosen with a very weak control field $\Omega_2=0.01\Gamma$ in (a)-(d).\ The sharp transparency window on top of the large absorption width appears when temperature is lower, which signifies a reduction of momentum distribution.\ In (e), the group velocity behaves similarly to Fig. \ref{counter_B}(e) at the low $T$ limit which saturates as atomic motions do not matter anymore, while at higher $T$ we still see slow light up to the temperature of mK (not shown) though EIT is very weak due to the weak control field we choose here.\ The relatively large $\Gamma$ of D2 transition makes the decoherence of Doppler effects possible only when  $T\gtrsim\hbar\Gamma/k_B$.\

\subsubsection{Fermions}

\begin{figure}[t]
\centering\includegraphics[height=6.0cm, width=8.5cm]{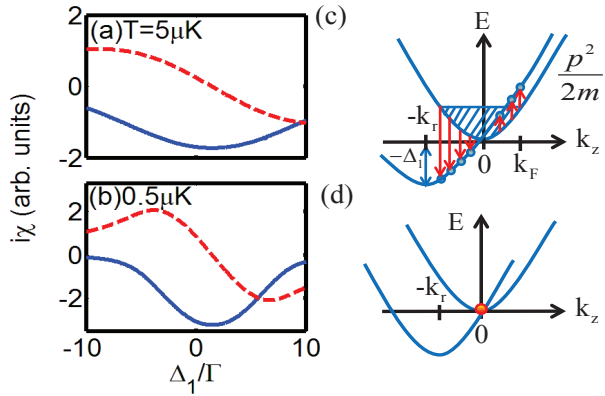}
\caption{(Color online) Counterpropagating excitation scheme for EIT in fermions.\ (a,b) Absorption (solid-blue) and dispersion (dash-red) profiles are plotted for $^{40}$K at temperatures of $5$ and $0.5~\mu$K (Fermi temperature $T_f=1.98\mu$K).\ (c) Two energy bands of free and shifted-recoil particles are shown as $p^2/(2m)$ and $-\Delta_1+(p+p_r)^2/(2m)$ respectively ($\Delta_1$ is chosen to be equal to recoil energy for example).\ The energy difference between two energy bands is demonstrated for fermions where solid-red arrows represent different excitation paths from the Fermi surface.\ (d) Same energy bands of (c) are shown for bosons.\ The  zero momentum ($p=0$) state (red filled circle) match these two bands exactly for Bose-Einstein condensation.}%
\label{counter_F}
\end{figure}

For a noninteracting fermionic gas, it is straightforward to derive the EIT property by using fermionic number distribution in Eq. (\ref{quantum}).\ In Fig. \ref{counter_F}, we show the electric susceptibility of probe field in $^{40}$K atoms.\ The excited state is chosen as low-lying Rydberg states $|21\text{P}_{3/2}\rangle$ (transition wavelength $288$ nm) with spontaneous decay time $25.4$ $\mu$s.\ In Fig. \ref{counter_F} (a) and (b), we find no EIT across the Fermi temperature $T_f=1.98~\mu$K which indicates a mismatch of resonance condition for the probe field as shown in (c).\ This mismatch comes from the Fermi distribution of atoms that has a Fermi surface of momentum and the recoil energy shift, $E_r$.\ It contrasts with bosonic atoms that mostly distributes around zero momentum as shown in (d) when condensation sets in.\ The absorption peaks in (a) and (b) can also be identified by the recoil energy of probe laser, $\Delta_1=E_1/\hbar$.\ The effect of the temperature is the absorption width that is larger for a higher temperature.\ Similar to Rydberg bosons, we may increase $\Omega_2$ to overcome the momentum mismatch that EIT emerges.
For D2 transitions, we have similar results as bosons in Fig. \ref{D2} (a,b) where the transparency window shows up on top of the large absorption profile when a weaker control field is used ($\Omega_2\ll\Gamma$).\ As the control field increases, e.g. $\Omega_2\approx\Gamma$, the EIT profile does not change across $T_f$ as if atomic motions are frozen.

\subsection{Results of copropagating excitation scheme}

Here we consider the copropagating excitation scheme where the recoil momentum is reduced to the energy transfer of hyperfine splitting of the ground states which are $6.834$ and $1.285$ GHz for $^{87}$Rb and $^{40}$K respectively.\ The main difference between co- and counter-propagating excitation schemes is the magnitude of recoil energy ($E_r$) that decides the transparency condition.\ From Eqs. (\ref{quantum}) and (\ref{kernel}),we can see that when $\p_r\approx 0$, $F_p$ is almost independent of $\p$, so that the temperature dependence of the EIT effects will not be expected significantly.\

\begin{figure}[t]
\centering\includegraphics[height=4.6cm, width=8.5cm]{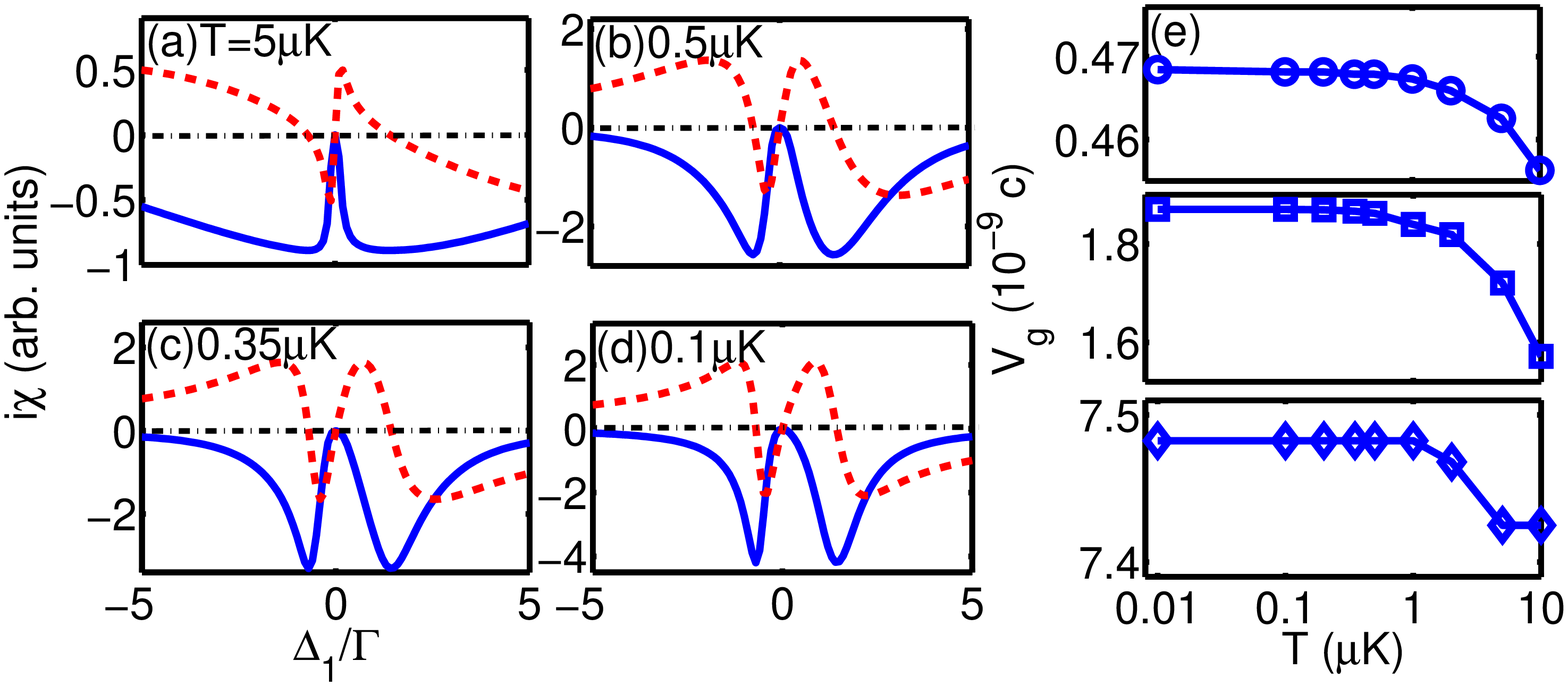}
\caption{(Color online) Copropagating excitation scheme for EIT in bosons.\ Absorption (solid-blue) and dispersion (dash-red) profiles of $^{87}$Rb are shown in the same temperature conditions (a)-(d) as in Fig. \ref{counter_B}.\ In (e), the group velocity $v_g$ is plotted for various control fields, $\Omega_2/\Gamma=0.5(\bigcirc)$, $1(\Box)$, $2(\Diamond)$ over temperatures from $0.01$ to $10\mu$K in log scale, which shows significant reduction and saturates in the low $T$ limit.\ High temperature limit has more reduction due to the narrower transmission window.\ The solid lines are drawn to connect data points.}%
\label{co_B}
\end{figure}

First we consider low-lying Rydberg bosons and study the temperature dependence of EIT in Fig. \ref{co_B}.\ When $T<T_c$ as shown in Fig. \ref{co_B}(c) and (d), it is similar to the counterpropagating scheme where the probe field interacts largely with condensation particles that have zero momentum distribution.\ There are again two absorption peaks similarly identified as Rydberg bosons in the counterpropagating excitation scheme with $E_r\approx 0$, and the transparency condition is $\Delta_1\approx 0$ due to negligible $E_r$.\ When $T\gg T_c$, the absorption widths are broadened due to atomic motions, which makes the transmission window narrower as in (a).\ In (e) we calculate the group velocity at the transparency condition.\ It is plotted over a range of temperature from $0.01$ to $10$ $\mu$K in log scale for various control fields from $0.5$ to $2\Gamma$.\ It saturates in the low $T$ limit, the same as the counterpropagating scheme, but becomes smaller at higher $T$ due to the narrower transmission window.\

\begin{figure}[t]
\centering\includegraphics[height=4.6cm, width=8.5cm]{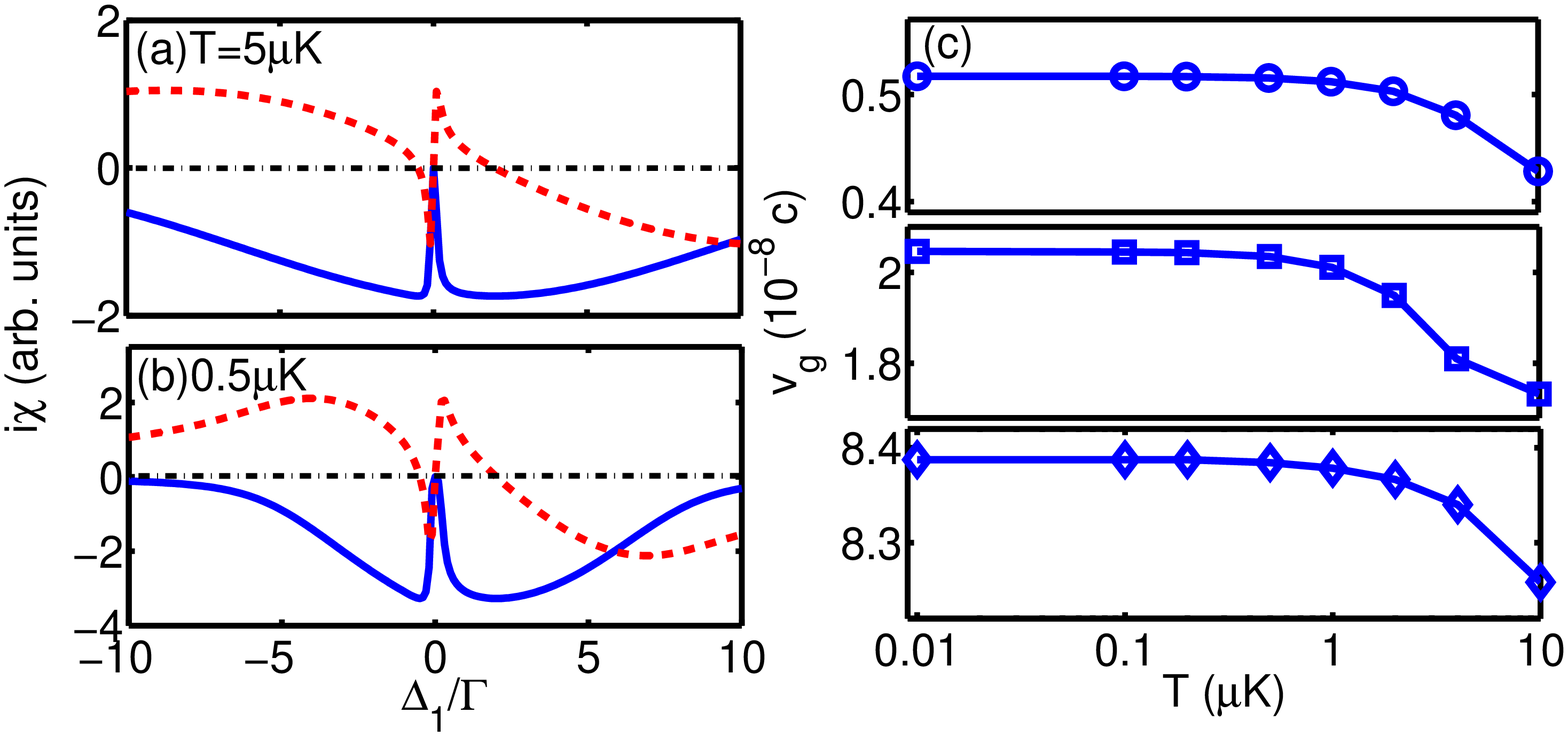}
\caption{(Color online) Copropagating excitation scheme for EIT in fermions.\ Absorption (solid-blue) and dispersion (dash-red) profiles of $^{40}$K.  Same temperature conditions are used as in Fig. \ref{counter_F}.\ Group velocity $v_g$ is plotted for temperatures from $0.01$ to $10\mu$K in log scale in (c) for various control fields, $\Omega_2/\Gamma=0.5(\bigcirc)$, $1(\Box)$, $2(\Diamond)$.\ It saturates in the low $T$ limit and more reduction of velocity is observed due to the narrower transmission window at higher $T$.\ The solid lines are drawn to connect data points.}
\label{co_F}
\end{figure}

In Fig. \ref{co_F}, we consider low-lying Rydberg fermions, and investigate the EIT property and group velocity.\ Absorption peaks can be identified in the same way as bosons in copropagating scheme.\ For lower temperature in (b), the widths of absorption peaks are narrower indicating a momentum distribution near the Fermi surface.\ They can be estimated in the order of $2\sqrt{E_F E_1}/\hbar$ from Eq. (\ref{kernel}) where $E_F$ is the Fermi energy.\ The separation between the peaks approaches $2\Omega_2$ when the control field is stronger.\ The transmission window becomes narrower when $T$ is higher, which decreases the group velocity as shown in Fig. \ref{co_F}(c).\ Contrary to the counterpropagating scheme, there is always a transparency window for the temperatures considered here with $\Omega_2=\Gamma$.\ It is due to negligible recoil energy shift that two-photon resonance condition is reached even for large atomic momentum distribution at high temperature.\

For D2 transitions with a larger $\Gamma$, similar to the Rydberg bosons/fermions, there is a transparency window due to negligible recoil energy shift for $\Omega_2$ down to $0.001\Gamma$.\ And again, the EIT appears on top of the absorption profile as in the counterpropagating scheme with D2 transitions.\

\section{Weakly interacting Bose-Einstein condensate}

Here we include the interaction energy in the noninteracting Hamiltonian of Bose gas and investigate the EIT in a weakly interacting Bose-Einstein condensate (BEC).\ The interaction Hamiltonian is
\begin{eqnarray}
H_{U}&=&\frac{1}{2V}\sum_{\p,\p^{\prime},\q}\left[U_{aa}\hat{a}_{\p+\q}^{\dag}\hat{a}_{\p^{\prime}-\q}^{\dag}\hat{a}_{\p^{\prime}}\hat{a}_{\p}\right.
\nonumber\\&+&\left.U_{bb}\hat{b}_{\p+\q}^{\dag}\hat{b}_{\p^{\prime}-\q}^{\dag}\hat{b}_{\p^{\prime}}\hat{b}_{\p}
+U_{ee}\hat{e}_{\p+\q}^{\dag}\hat{e}_{\p^{\prime}-\q}^{\dag}\hat{e}_{\p^{\prime}}\hat{e}_{\p}\right]\nonumber\\
&+&\frac{1}{2V}\sum_{\p,\p^{\prime},\q}\left[ 2U_{ab}\hat{b}_{\p+\q}^{\dag}\hat{a}_{\p^{\prime}-\q}^{\dag}\hat{a}_{\p^{\prime}}\hat{b}_{\p}\right.
\nonumber\\&+&\left.2U_{ae}\hat{e}_{\p+\q}^{\dag}\hat{a}_{\p^{\prime}-\q}^{\dag}\hat{a}_{\p^{\prime}}\hat{e}_{\p}+
2U_{be}\hat{e}_{\p+\q}^{\dag}\hat{b}_{\p^{\prime}-\q}^{\dag}\hat{b}_{\p^{\prime}}\hat{e}_{\p}\right],\nonumber\\
&\approx&\frac{U_{aa}}{2V}\sum_{\p,\p^{\prime},\q}\hat{a}_{\p+\q}^{\dag}\hat{a}_{\p^{\prime}-\q}^{\dag}\hat{a}_{\p^{\prime}}\hat{a}_{\p}+\rho_cU_{ab}\sum_{\p}\hat{b}_{\p}^{\dag}\hat{b}_{\p}
\nonumber\\&+&\rho_cU_{ae}\sum_{\p}\hat{e}_{\p}^{\dag}\hat{e}_{\p},\label{interaction}
\end{eqnarray}
where in the first expression we show in general all the possible atom-atom interactions $U_{st}=4\pi\hbar^2 a_{st}/m$ with s-wave scattering lengths $a_{st}$ in our $\Lambda$ type atomic configuration ($s,t\equiv a,b$ or $e$).\ In the second expression we use the Hartree-Fock approximation \cite{BEC} for the interactions of $U_{ab}$, $U_{ae}$, and $\rho_c\equiv\langle\hat{a}_{0}\rangle^{2}/V$ is the mean field condensate density.\ The neglected terms include $U_{bb}$, $U_{ee}$, $U_{be}$, and other interactions between the atomic fields $\hat{b}_\p$, $\hat{e}_\p$, and $\hat{a}_{\p\neq 0}$.\ They are at least $\mathcal{O}(\Omega_{1}^{2})$ which are negligible in the weak field limit.\ We then absorb the mean field energy shifts of the bare atomic fields $\hat{b}_\p$ and $\hat{e}_\p$ into the noninteracting Hamiltonian of Eq. (\ref{nonH}), which becomes (cf. Eq. (\ref{M}))

\begin{eqnarray}
&&\hat{M}'_\p=\nonumber\\
&&
\begin{bmatrix}
\Delta_{1}+\frac{\p^{2}}{2m} & 0 & -\Omega_{1}\\
0 & \Delta_{2}+\frac{\left(  \p+\p_r\right)  ^{2}}{2m}+\rho_cU_{ab} & -\Omega_{2}\\
-\Omega_{1} & -\Omega_{2} & \frac{\left(  \p+\p_{1}\right)  ^{2}}{2m}+\rho_cU_{ae}%
\end{bmatrix},\nonumber\\
&&=\rho_cU_{ae}\hat{I}+\begin{bmatrix}
\tilde{\Delta}_{1}+\frac{\p^{2}}{2m} & 0 & -\Omega_{1}\\
0 & \tilde{\Delta}_{2}+\frac{\left(  \p+\p_r\right)  ^{2}}{2m}& -\Omega_{2}\\
-\Omega_{1} & -\Omega_{2} & \frac{\left(  \p+\p_{1}\right)  ^{2}}{2m}%
\end{bmatrix},\label{M_prime}
\end{eqnarray}
where we have extracted the common mean field energy $\rho_cU_{ae}$ that we derive the effective noninteracting Hamiltonian with shifted detunings $\tilde{\Delta}_1=\Delta_{1}-\rho_cU_{ae}$ and $\tilde{\Delta}_2=\Delta_{2}+\rho_c(U_{ab}-U_{ae})$ in the last line.\ 

Using the dark state $\hat{\beta}_\p$ introduced in Sec. 3 and with Eq. (\ref{M_prime}), we let $\hat{a}_\p\approx\hat{\beta}_\p$ and $\hat{\beta}_{\p}=\sqrt{N_c}
\delta_{\p=0}+\hat{\beta}_{\p\neq 0}$ where $\hat{\beta}_{\p\neq 0}$ are quasiparticles.\ We then apply Hartree-Fock-Bogoliubov and Popov approximations for the interactions of $U_{aa}$ in Eq. (\ref{interaction}), and diagonalize the total Hamiltonian by Bogoliubov transformation.\ Define $H_D=H_A+H_{AL}+H_U$ and introduce the chemical potential $\mu$ to take care of the total number conservation, we have
\begin{eqnarray}
&&H_{D}-\mu\hat{N}=E_g-\mu N_c+\sum_{\q\neq 0}\epsilon(\q)\alpha_\q^\dag\alpha_\q,\\ &&\text{where}\nonumber\\
&&E_g=\tilde{\epsilon}_{D}(\p=0)N_c+\frac{U_{aa}}{2V}(N_c^2+2N_{ex}^2)+\frac{N_c N U_{ae}}{V}\nonumber\\&&+\sum_{\q>0}[\epsilon-\epsilon_{1}(\q)]
\end{eqnarray}
is the ground state energy including the mean field energy and energy correction, and $\tilde{\epsilon}_{D}(\p)=\tilde{\Delta}_1+\p^2/(2m)$.\ The eigenvalue and eigenstates of the Hamiltonian are $\epsilon(\q)=(\epsilon_1^2(\q)-\epsilon_2^2)^{1/2}$ and $\alpha_\q=\text{cosh}\theta(\q)\hat{\beta}_{\q}+\text{sinh}\theta(\q)\hat{\beta}^\dag_{-\q}$ where $\text{tanh}2\theta(\q)=\epsilon_2/\epsilon_1(\q)$, $\epsilon_{1}(\q)=\tilde{\epsilon}_{D}(\q)-\mu+2\rho_cU_{aa}$, and $\epsilon_{2}=\rho_cU_{aa}$.\ The chemical potential is $\mu=\hbar\tilde{\Delta}_1+\rho_cU_{aa}$ which is found by regarding gapless Bogoliubov excitation energy, $\epsilon(\q=0)=0$.\

Using the obtained new ground state wavefunction, we derive the electric susceptibility for the probe field in the weakly interacting BEC,
\begin{eqnarray}
\chi(\omega_1)&=&\chi_{C}(\omega_1)+\chi_{QD}(\omega_1)+\chi_{TH}(\omega_1),\nonumber\\
&=&\frac{\rho d_{ae}^2}{\hbar}\Big[\frac{\rho_c}{\rho}\tilde{F}_{\p=0}+\frac{1}{N}\sum_{\p\neq0}\tilde{F}_\p\text{sinh}^2\theta\nonumber\\
&+&\frac{1}{N}\sum_{\p\neq0}\tilde{F}_\p n_b\text{cosh}2\theta\Big],
\end{eqnarray}
where we separate the condensate(C), quantum depletion (QD), and thermal depletion (TH) contributions to $\chi(\omega_1)$.\ The shifted kernel function is $\tilde{F}_\p$ with $\Delta_1\rightarrow\tilde{\Delta}_1$ and $\Delta_2\rightarrow\tilde{\Delta}_2$.\ Bose distribution is $n_b(\q)=1/[e^{\epsilon(\q)/(k_BT)}-1]$ for Bogoliubov quasi-particles \cite{BEC}.\

\begin{figure}[t]
\centering\includegraphics[height=4.6cm, width=8.5cm]{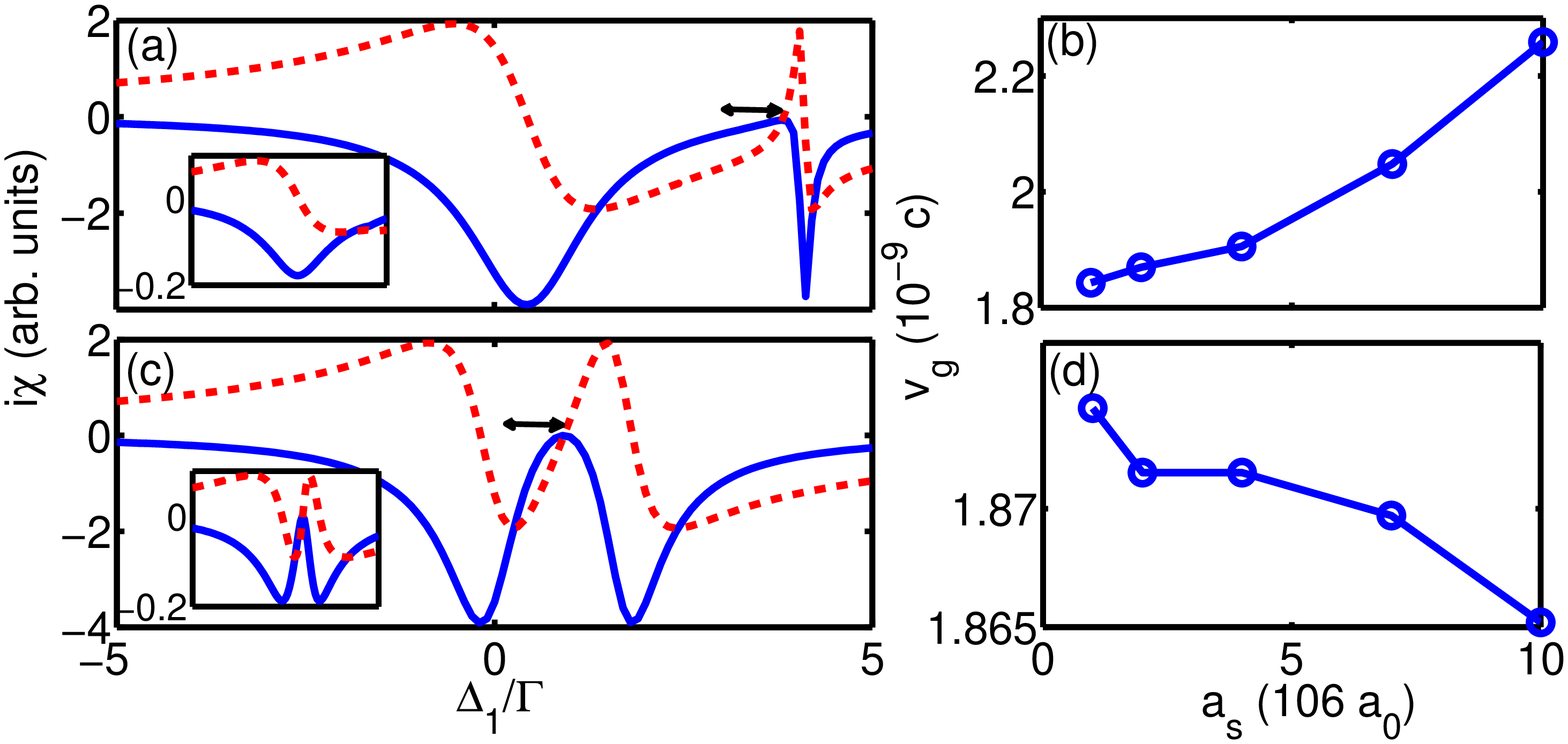}
\caption{(Color online) EIT property in weakly interacting $^{87}$Rb condensate.\ Dispersion (dash-red) and absorption (solid-blue) profiles are shown in (a) counter- and (c) copropagating excitation schemes.\ The respective group velocities (b) and (d) are calculated at the transparency conditions and plotted over interaction strengths ($a_s$).\ The scattering length and temperature are chosen as $a_s=7\times 106~a_0$ and $T=10$nK in (a) and (c) where the insets are $i(\chi_{QD}+\chi_{TH})$ from Bogoliubov particles $\rho_{ex}\approx 0.1\rho$.\ The coupling atomic levels and the atomic density $\rho$ are the same as in Fig. \ref{counter_B}, and the double arrows in (a) and (c) are the mean field interaction energy shifts, $\rho_c U_{aa}$.\ In (b) and (d), we draw the solid lines for eye-guiding.}
\label{bogo2}
\end{figure}

Without loss of generality, we let $U_{ab}=U_{aa}$, and assume $U_{ae}\approx 0$ (which is just to have an excitation energy shift).\ In Fig. \ref{bogo2}, we demonstrate the EIT property for counter- and co-propagating excitation fields in weakly interacting Bose-Einstein condensates using low-lying Rydberg transitions.\ We choose the scattering length $a_s$ seven times of $106a_0$ for $^{87}$Rb where $a_0$ is Bohr radius, and explore the effects of quantum depletion ($\rho_{ex}\approx 0.1\rho$ with negligible thermal depletion at $T=10$nK) on the profiles.\ The overall profiles in (a) and (c) bear the main features for noninteracting Bose gas below $T_c$ in Fig. \ref{counter_B}(d) and \ref{co_B}(d) respectively.\ The mean field interaction energy, $\rho_c U_{aa}\approx 0.9\Gamma$, shifts the transparency conditions of noninteracting BEC, which we denote as the double arrows in (a) and (c).\ It is of kHz that can be observable in experiments.\ Two absorption peaks are still resolved and the transparency conditions are mainly determined by the condensate particles.\ The insets from the Bogoliubov excitations can be seen similarly as in Fig. \ref{counter_B}(b) and \ref{co_B}(b) respectively at temperature above $T_c$.\ $\chi_{QD}(\omega_1)$ and $\chi_{TH}(\omega_1)$ are contributions from finite momenta which broaden the shapes of absorption peaks.\ This interaction-induced broadening in interacting Bose-Einstein condensates can be observed through increasing the interaction energy via the Feshbach resonance.\ In Fig. \ref{bogo2}(b) and (d), we calculate the group velocity near the transparency point and find that it decreases (increases) for co(counter)-propagating excitation scheme when the interaction strength increases.\ It is due to the narrower (broader) transmission window from the Bogoliubov particle contributions.\ For even stronger interaction strengths, the Bogoliubov approximation fails.

\section{Conclusion}
We theoretically investigate the electromagnetic induced transparency in ultracold atomic gases.\ The atomic dynamics and quantum statistics significantly modify the light propagating properties in quantum degenerate gases concerning noninteracting bosons, fermions, and weakly interacting Bose-Einstein condensate for co- and counter-propagating excitation schemes.\ Throughout the systematic studies of EIT property, we carefully analyze the interplay of three energy scales: control field strength, temperature, and spontaneous emission rate.\  We use low-lying Rydberg excited states ($|21\text{P}\rangle$ of K and $|24\text{P}\rangle$ of Rb as examples) and D2 transitions to demonstrate the EIT property across the characteristic temperatures ($T_c,T_f$) of noninteracting bosons and fermions.\ In the counter-propagating scheme, we show a transition of the EIT property across $T_c$ of bosons when the control field is close to the recoil energy.\  It can be used as a sensitive determination of $T_c$.\  For fermions at low temperature, the presence of the Fermi sea can destroy the EIT effect, and the absorption widths can be determined by the Fermi energy.\  As the control field increases, the EIT profile can be described by the classical theory as if atomic motions are frozen.\  The results of D2 transitions are similar to those of the Rydberg states.\  In the copropagating scheme, EIT is robust for both bosons and fermions in a wide range of temperature even with a weak control field due to negligible recoil energy.\ We note that the laser frequency for low-lying Rydberg transitions is more challenging than for D2 transitions, but it can be generated through frequency doubling of a pulse-amplified cw laser \cite{30P,laserF}.\ We may use lower low-lying Rydberg transitions for less demanding laser frequency, and we expect a similar transition of the EIT property as long as the control field strength and spontaneous emission rate are in the order of the recoil energy.\ Note that for such a low control field strength in the EIT experiment, the dephasing rate of the ground states may influence the transmission of the probe field.\ The recent progress of creating a coherent optical memory by minimizing the inelastic collisions of the Bose-Einstein condensate has reached the coherence time much larger than 1 ms \cite{coherence_Hau}.\ Therefore, we expect the ground state dephasing is negligible in our results.

We further apply the Hartree-Fock-Bogoliubov mean field approach to the weakly interacting Bose-Einstein condensate, and find that EIT is modified by quantum and thermal depletions due to excitation particles.\ The mean field energy from atom-atom interactions shifts the transparency condition, and an interaction induced broadening is shown in absorption widths which can be observable by tuning atomic scattering lengths via Feshbach resonance.\ The group velocity is also calculated and analyzed in details for various schemes of different temperatures, control field or interaction strengths.\ Future developments might involve using EIT to investigate the many-body and strongly interacting atomic system, and it may provide a fruitful information of atomic dynamics or synthesize magnetic gauge fields \cite{gauge1} to simulate spin-orbit coupling quantum degenerate gases \cite{gauge2}.  
\section*{ACKNOWLEDGMENTS}
We acknowledge the support by NSC, Taiwan, R. O. C.\ IAY is the partner in EU FP7 IRSES project COLIMA (contract PIRSES-GA-2009-247475) and acknowledges the support from National Tsing Hua University 101N2713E1 grant.

\end{document}